\documentclass[sigconf]{acmart} 

\newcommand{\ignore}[1]{}
\AtBeginDocument{%
  \providecommand\BibTeX{{%
    \normalfont B\kern-0.5em{\scshape i\kern-0.25em b}\kern-0.8em\TeX}}}

\copyrightyear{2025}
\acmYear{2025}
\setcopyright{rightsretained}
\acmConference[Koli Calling '25]{Proceedings}{Date}{Koli, Finland}
\acmBooktitle{25th Koli Calling International Conference on Computing Education Research}
\acmDOI{XXXX}
\acmISBN{XXX}

\AtBeginDocument{  \providecommand\BibTeX{{
  Bib\TeX}}}

\makeatletter

\gdef\@copyrightpermission{
  \begin{minipage}{0.3\columnwidth}
   \href{https://creativecommons.org/licenses/by/4.0/}{\includegraphics[width=0.90\textwidth]{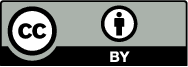}}
  \end{minipage}\hfill
  \begin{minipage}{0.7\columnwidth}
   \href{https://creativecommons.org/licenses/by/4.0/}{This work is licensed under a Creative Commons Attribution International 4.0 License.}
  \end{minipage}
  \vspace{5pt}
}
\makeatother

\usepackage{enumitem}
\usepackage{comment}
\usepackage{soul}
\setlist[itemize]{noitemsep, topsep=0pt}
\setlist[enumerate]{noitemsep, topsep=0pt}
\usepackage{graphicx, array, booktabs, makecell, multirow, subfig}
\newcolumntype{L}[1]{>{\raggedright\let\newline\\\arraybackslash\hspace{0pt}}m{#1}}
\newcolumntype{C}[1]{>{\centering\let\newline\\\arraybackslash\hspace{0pt}}m{#1}}
\newcolumntype{R}[1]{>{\raggedleft\let\newline\\\arraybackslash\hspace{0pt}}m{#1}}

\begin{document}
%TC:macro \cite [option:text,text]
%TC:macro \citep [option:text,text]
%TC:macro \citet [option:text,text]
%TC:envir table 0 1
%TC:envir table* 0 1
%TC:envir tabular [ignore] word
%TC:envir displaymath 0 word
%TC:envir math 0 word
%TC:envir comment 0 0
\title{Evolution of Programmers' Trust in Generative AI Programming Assistants}

%%TODO Authors
\author{Anshul Shah}
\email{ayshah@ucsd.edu}
\affiliation{%
  \institution{University of California, San Diego}
  \country{USA}
}

\author{Thomas Rexin}
\email{tjrexin@ncsu.edu}
\affiliation{%
  \institution{North Carolina State University}
  \country{USA}
}

\author{Elena Tomson}
\email{etomson@ucsd.edu}
\affiliation{%
  \institution{University of California, San Diego}
  \country{USA}
}

\author{Leo Porter}
\email{leporter@ucsd.edu}
\affiliation{%
  \institution{University of California, San Diego}
  \country{USA}
}

\author{William G. Griswold}
\email{bgriswold@ucsd.edu}
\affiliation{%
  \institution{University of California, San Diego}
  \country{USA}
}

\author{Adalbert Gerald Soosai Raj}
\email{asoosairaj@ucsd.edu}
\affiliation{%
  \institution{University of California, San Diego}
  \country{USA}
  }



\begin{abstract}
\textbf{Motivation.} Trust in generative AI programming assistants is a vital attitude that impacts how programmers use those programming assistants. Programmers that are \textit{over-trusting} may be too reliant on their tools, leading to incorrect or vulnerable code; programmers that are \textit{under-trusting} may avoid using tools that can improve their productivity and well-being. 

\textbf{Methods.} Since trust is a dynamic attitude that may change over time, this study aims to understand programmers' evolution of trust after immediate (one hour) and extended (10 days) use of GitHub Copilot. We collected survey data from 71 upper-division computer science students working on a legacy code base, representing a population that is about to enter the workforce. Leveraging existing survey instruments and open-ended free response questions, we quantitatively measure student trust levels and qualitatively uncover why student trust changes.

\textbf{Findings.} Student trust, on average, increased throughout the study. After completing a project with Copilot, however, students felt that Copilot requires a competent programmer to complete some tasks manually. Students mentioned that seeing Copilot's correctness, understanding how Copilot uses context from the code base, and learning some basics of natural language processing contributed to their elevated trust. On the other hand, students pointed to Copilot's incorrect code suggestions as one of their main reasons for having less trust in the tool.

\textbf{Implications.} Our study helps instructors and industry managers understand the factors that influence how students \textit{calibrate their trust} with generative AI programming assistants. We make four pedagogical recommendations, which are that CS educators should 1) provide opportunities for students to work with Copilot on challenging software engineering tasks to calibrate their trust, 2) teach traditional skills of comprehending, debugging, and testing so students can verify output, 3) teach students about the basics of natural language processing, and 4) explicitly introduce and demonstrate the range of features available in Copilot.
%i think your slack is muted lol, just gonna keep putting comments

\end{abstract}





\maketitle

\section{Introduction}
\label{sec:intro}

Adoption of generative AI assistants is growing among the software engineering industry, with GitHub Copilot in particular being the most widely-adopted AI tool among businesses \cite{copilot_adoption}. Whether our students join start-up companies or established tech companies, it is highly likely that they will use generative AI programming assistants in their work \cite{Mehta_2025, Langley_2024}. Trust in these AI programming assistants is a vital attitude for educators and researchers to consider, since a programmers' trust in an AI tool impacts their reliance on the tool \cite{Liao_MATCH, Wang_TrustInCopilot}. Developers who are over-trusting of their AI assistants may introduce insecure or incorrect code \cite{Perry_InsecureCode, Prather_Weird, Amoozadeh_Trust}, while developers who are under-trusting may miss out on benefits of AI assistants, such as improved productivity and developer well-being \cite{Shihab_BrownfieldCopilot, Cui2024Productivity, peng2023impact, Github_survey}. 

Our study aims to understand \textit{how} and \textit{why} student trust changes as they learn about and use GitHub Copilot. While the majority of research on student attitudes and use of generative AI assistants has focused on introductory computer science courses (CS) \cite{Prather_Weird, Vadaparty_CS1LLM, Denny_Era, Prather_WideningGap, Amoozadeh_Trust}, we focus on upper-division CS students' trust in AI assistants while working on an existing code base. Our findings aim to inform computing educators how to foster students' AI literacy and calibrate students' trust levels to prepare students for a software engineering field that is embracing generative AI assistants. \citeauthor{Ozkaya_Motivation} summarizes the importance of adapting our instruction to align with the reality of the software engineering industry:

%\begin{quote}
    \textit{``Software engineering and computer science education has already missed the boat by continuing to focus on teaching green field development while today the reality of system development is brownfield... We teach students hello world development, while we should be teaching them how to read millions of lines of code, triage and fix bugs that they have not contributed to... With LLMs and their sister AI-driven apps assisting developers, we need to be teaching next-generation software engineers when to trust, how to create evidence to trust, how to do trust assessment rapidly and correctly, and how to improve such assistants.''} \cite{Ozkaya_Motivation}
%\end{quote}

To lay the groundwork for computing educators to teach students how to calibrate their trust with AI assistants while working on large code bases (i.e., brownfield development), we ask the following two research questions:

\begin{description}
    \item[RQ1:] How does student trust in GitHub Copilot change after \textit{immediate} and \textit{extended} use of GitHub Copilot in a large code base?
    \item[RQ2:] What factors contribute to students' evolution of trust in GitHub Copilot in a large code base?
\end{description}

We studied how 71 upper-division CS students' trust levels change 1) after attending a lecture in which they learned basics about natural language processing, saw demonstrations of  features that are useful for brownfield programming, and worked with Copilot for a small task and 2) after completing a 10-day project where they added a feature to a large code base using GitHub Copilot. Using a survey to collect repeated measures of student trust and open-ended questions to understand \textit{why} students' trust changes, we found that, on average, trust in GitHub Copilot increased after immediate and extended use. Students' \textit{immediate} trust attitudes were influenced by Copilot's code generation abilities, Copilot's context-awareness features that support program comprehension, and understanding Copilot's inner-workings. After \textit{extended} Copilot use, however, students overwhelmingly highlighted the importance of having a ``competent programmer'' to verify or correct Copilot's output. Our findings lead to four concrete recommendations for how CS instructors can teach students how to effectively and responsibly use AI assistants.

% Its important to understand how they are creating their trust perceptions to be able to know how to address issues with trust.

% If we can also show times when trust is incorrect, they can know when to verify. Not only knowing how it works, but also knowing how they as humans should work with the tools.

% Understanding how they make their decisions to use/not use or trust vs not trust.

\section{Theoretical Framework: The MATCH Model}
\label{sec:theory}

\begin{figure*}
    \centering
    \includegraphics[scale=0.48]{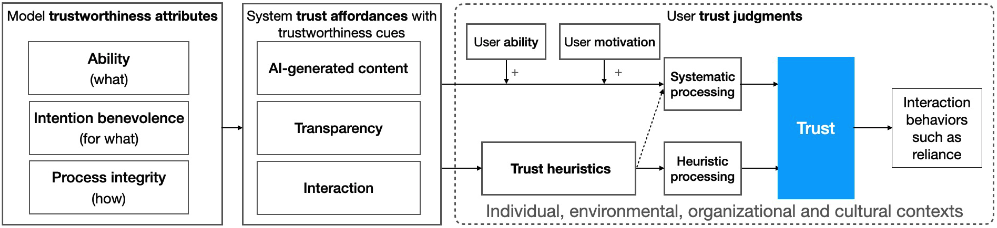}
    \caption{The MATCH model, reproduced exactly as presented by \citeauthor{Liao_MATCH} \cite{Liao_MATCH}.}
    \label{fig:match_model}
\end{figure*} 

Early models of trust formation emerged from social science research. The Ability-Benevolence-Integrity (ABI) model, developed in 1995 to explain interpersonal trust formation \cite{ABI_Model}, posits that the three characteristics of a trustee that impact trust are: 1) the trustee's ability (i.e., competency) 2) the trustee' benevolence (i.e., aligned intentions and goals between), and 3) the trustee's integrity (i.e., adhering to a clear set of principles) \cite{ABI_Model}. In a similar vein, \citeauthor{Lee_Moray_Automation}'s model of trust formation between humans and automation says that trust depends on a system's performance (i.e., ability), purpose (i.e., benevolence), and process (i.e., integrity) \cite{Lee_Moray_Automation}. 

One limitation of these models, however, is that they only describe characteristics of the trustee, or automated system. As a result, \citeauthor{sundar2008_affordances} presented the MAIN model in 2008 that highlighted the importance of \textit{system affordances} for trust formation between humans and technology \cite{sundar2008_affordances}. Broadly, affordances are properties of a system that act as ``cues'' to communicate how a user can engage with or act upon the system \cite{sundar2008_affordances}. For example, an empty input box---an Interactivity affordance---signifies to a user that they may type a response in that open space \cite{sundar2008_affordances}. \citeauthor{sundar2008_affordances} notes that system affordances play a prominent role in shaping a human's trust \cite{sundar2008_affordances}, expanding previous models of trust that focused only on a system's attributes.

The MATCH model, presented by \citeauthor{Liao_MATCH} in 2022, describes the trust formation process specifically for human-AI interaction \cite{Liao_MATCH}. The MATCH model, reproduced in Figure \ref{fig:match_model}, explains trust formation as an expression of a model's underlying \textit{trustworthiness attributes} that are conveyed to a user via \textit{trust affordances}, which are processed by the user as \textit{trust judgments} \cite{Liao_MATCH}. Thus, the model captures the entire chain of the human-AI interaction loop---from the model's true characteristics to the user's judgment of the model's affordances. The following sections describe and operationalize the three key components of trustworthiness attributes, trust affordances, and trust judgments in the context of programming with GitHub Copilot.

\subsection{Trustworthiness Attributes}

Trustworthiness attributes are the underlying characteristics of a model that are not directly observable by the user \cite{ABI_Model, Liao_MATCH}. Importantly, these attributes are \textit{not} the same as the model's output, which is a \textit{result} of the model's underlying attributes. \textit{Ability} refers to Copilot's capabilities in comprehending and generating code. \textit{Intention Benevolence} refers to the alignment of goals and desires between Copilot and the programmer. \textit{Process Integrity} refers to the underlying large language model architecture that is responsible for Copilot's text and code generation.

% \begin{itemize}
%     \item \textit{Ability} refers to Copilot's capabilities in comprehending and generating code.
%     \item \textit{Intention Benevolence} refers to the alignment of goals and desires between Copilot and the programmer.
%     \item \textit{Process Integrity} refers to the underlying large language model architecture that is responsible for Copilot's text and code generation.
% \end{itemize}

\subsection{Trust Affordances}

Trust affordances are features that are designed to communicate and convey the underlying trustworthiness attributes to a user \cite{Liao_MATCH, sundar2008_affordances}. Affordances act as a link between the underlying trustworthiness attributes and user trust judgments. \textit{AI-generated content} refers to the output from Copilot, whether it is generated code, text descriptions of the code, documentation, etc. \textit{Transparency} refers to features that give users insight into how Copilot generated its output, such as by showing the files and code snippets it used as additional context when a user submits a prompt to Copilot chat. \textit{Interaction} refers to the ways in which a programmer can interact with Copilot and, more broadly, the user options that are available \cite{Liao_MATCH}. In Copilot, some of the interaction features include in-line code generation, in-line chat, a chat window to the side, and code generation via comments. However, programmers also have control over the prompt, including the ability to add context (i.e., files or directories in the code base) to prompts, add certain keywords to prompts to ask specific questions, etc.

\subsection{Trust Judgments}

Trust judgments refer to how a user, or programmer, processes the output of a model to form their trust attitudes \cite{Liao_MATCH}. There are two ways in which programmers may make trust judgments: via systematic processing or heuristic processing. \textit{Systematic processing} occurs when the user of an AI model rigorously evaluates the output to make a rational judgment on its trustworthiness \cite{Liao_MATCH}. In Copilot, the output can be either generated code or text-based descriptions. Systematic processing, therefore, entails careful evaluation of the correctness and appropriateness of Copilot's output. \textit{Heuristic processing}, by contrast, involves the user or programmer using known heuristics, or general rules, to make quicker and easier judgments \cite{Liao_MATCH}. \citeauthor{Liao_MATCH} note that when users apply these heuristics, they are more prone to errors in judgment \cite{Liao_MATCH}.

% \begin{itemize}
%     \item Systematic Processing occurs when the user of an AI model rigorously evaluates the output to make a rational judgment on its trustworthiness \cite{Liao_MATCH}. In Copilot, 
%     \item Heuristic Processing, by contrast, involves the user or programmer using known heuristics, or general rules, to make quicker and easier judgments \cite{Liao_MATCH}. \citeauthor{Liao_MATCH} note that when users apply these heuristics, they are prone to errors in judgment \cite{Liao_MATCH}.
% \end{itemize}

\section{Related Work}
\label{sec:background}

\subsection{Trust in AI Assistants Among Experienced Developers}

Trust in AI assistants among software developers is vital for researchers to understand, as trust mediates how programmers use, or misuse, those tools. \citeauthor{Perry_InsecureCode} showed via an experimental study that participants who used an AI assistant introduced significantly more insecure code than participants who did not use the AI assistant, while also feeling a higher confidence that their code was secure \cite{Perry_InsecureCode}. \citeauthor{Perry_InsecureCode}'s study highlights the danger of an over-trusting developer, whereas developers who are under-trusting may not adopt AI assistants into their workflow \cite{Johnson_PICSE, Amoozadeh_Trust}, causing them to miss out on benefits of AI tool use that other studies have found, such as improved productivity \cite{Cui2024Productivity, peng2023impact, Github_survey, Shihab_BrownfieldCopilot} and reduced developer frustration \cite{Github_survey}. \citeauthor{Brown_Experience} highlights how early-career developers had higher acceptance rates of AI-suggested code than their senior peers \cite{Brown_Experience}, potentially indicating a greater level of trust in AI tools among newer developers.

Research has also uncovered what factors contribute to developers' trust. \citeauthor{Cheng_OnlineCommunities} outlines how online communities shape a developer's trust in AI assistants, using the MATCH model as a framework \cite{Cheng_OnlineCommunities}. Factors such as seeing clear demonstrations of the tool's features, how the tool performs in an authentic task with  ``lots of files'' and ``lots of context'', learning about the tool's decision-making process, and understanding the broader impacts of AI-generated code all contributed to developers' trust attitudes \cite{Cheng_OnlineCommunities}. Focusing on trust affordances, \citeauthor{Wang_TrustInCopilot} conducted interviews with developers to make design recommendations to foster greater trust in Copilot \cite{Wang_TrustInCopilot}. The authors pointed out trust affordances that Copilot's code generation features lacked, such as not showing its confidence in particular suggestions or whether other developers accepted a suggestion \cite{Wang_TrustInCopilot}.  \citeauthor{Wang_TrustInCopilot}'s study, however, did not use a version of Copilot that had Copilot chat, which is used extensively by programmers to generate code \cite{Shihab_BrownfieldCopilot, Shah_Copilot}.

In the CS education research space, limited work has studied upper-division students' trust. While \citeauthor{Shihab_BrownfieldCopilot} focus on programmers' productivity and correctness, they do not analyze trust-related measures. \citeauthor{Amoozadeh_Trust} created a survey to measure students' trust in AI assistants and showed that students in an operating systems course (an upper-division course) exhibited lower trust than students in a CS2 course \cite{Amoozadeh_Trust}. Most similar to our analysis comes from \citeauthor{Shah_Copilot}, who uses the same instrument as \citeauthor{Amoozadeh_Trust} to analyze students' trust on both Copilot's code generation features \textit{and} its code comprehension features (e.g., chat prompts, ``/explain'' command, etc.) in the context of working with large code bases \cite{Shah_Copilot}. Students demonstrated significantly more trust in the code comprehension features than the code generation features \cite{Shah_Copilot}, though the authors did not include any qualitative analysis to understand why those trust attitudes formed. 

Altogether, our study aims to contribute to the research on how programmers use and trust AI assistants by 1) measuring the \textit{change in trust} among programmers who are having their first interactions with GitHub Copilot and 2) understanding \textit{why} students' trust changes so that we can target ways to regulate their trust in AI assistants.

\subsection{Trust in AI Assistants Among Novice Programmers}

Several studies have used interviews or observational studies to understand perceptions of generative AI assistants \cite{Hou_ErodingInteractions, Zastudil_Perspectives, Prather_Weird, Prather_WideningGap, Lau_BanIt}. In these interviews, instructors and students generally express distrust in these tools \cite{Hou_ErodingInteractions, Zastudil_Perspectives, Lau_BanIt}. For example, \citeauthor{Zastudil_Perspectives} heard from a majority of instructors (5 out of 6) and students (7 out of 12) that they are concerned about the trustworthiness of these tools \cite{Zastudil_Perspectives}. In contrast, \citeauthor{Hou_ErodingInteractions} notes that students prefer to use tools like ChatGPT for help-seeking, replacing traditional help-seeking resources such as classmates or office hours \cite{Hou_ErodingInteractions}, highlighting the extent of students' trust in these tools. 

Limited work has explicitly aimed to measure and understand novice programmers' \textit{trust} besides \citeauthor{Amoozadeh_Trust} \cite{Amoozadeh_Trust}. In their study, \citeauthor{Amoozadeh_Trust} adapted an existing survey instrument related to trust in automation and administered the survey to students in the USA and India to broadly understand students' trust in AI assistants, including ChatGPT and Copilot \cite{Amoozadeh_Trust}. Via an open-ended text response question, the authors noted that nearly half of the students expressed distrust in their response, while others espoused Copilot for being overall helpful and contributing to their learning \cite{Amoozadeh_Trust}. \citeauthor{Amoozadeh_Trust}'s finding, however, of lower-division students exhibiting higher trust than upper-division students highlights the concern that motivated this paper: students with uncalibrated trust assessments of AI assistants may exhibit ineffective or counterproductive use of those tools. 

\subsection{Measuring Trust}

We observe that there are three common ways to understand trust in the aforementioned literature: 1) survey instruments \cite{Amoozadeh_Trust, Shah_Copilot, Vereschak_HAI}, 2) interviews or written free-responses from developers \cite{Cheng_OnlineCommunities, Wang_TrustInCopilot, Liao_MATCH, Amoozadeh_Trust, Johnson_PICSE}, and 3) acceptance rate of AI-generated code \cite{Gao_2024, Prather_WideningGap, Github_survey}. 

Our study uses methods 1) and 2) to measure trust. We use the trust survey presented by \citeauthor{Amoozadeh_Trust} that is a modified version of an earlier survey related to trust in automation \cite{körber_2018}. We do not use acceptance rate to measure trust since acceptance rate is a \textit{behavior}, which is different from treating trust as an \textit{attitude} \cite{baltes2025_rethinking}. Further, \citeauthor{Shah_Copilot} showed that programmers preferred to use Copilot chat to generate code suggestions \cite{Shah_Copilot}, which makes calculating acceptance rates more difficult than simply determining whether the in-line code suggestion was accepted by the programmer. Regardless of the way in which trust is measured, \citeauthor{baltes2025_rethinking} notes that trust in AI literature should leverage trust frameworks to help synthesize findings among studies \cite{baltes2025_rethinking}. Therefore, we aim to contribute to the broader literature on how developers trust AI assistants by discussing our findings in terms of the MATCH model.

\section{Study Context}
\label{sec:context}

\subsection{Course Context}

This study was conducted in an advanced software engineering course that focuses on brownfield software development at UC San Diego---a large, public university in North America. The course is an elective that can only be taken after students complete the core software engineering class in which students learn about the software development process, including AGILE workflow, and create a web or mobile application from scratch with a team. At a high-level, the course covers topics such as program comprehension, project management, and generative AI programming assistants while working on large code bases. 

%The seven learning objectives outlined in our syllabus are displayed below.

% \textit{By the end of the term, students will be able to:}
% \begin{itemize}
%     \item \textit{Setup a development environment to build and work on software tools from source.}
%     \item \textit{Read, understand, and modify parts of a code base.}
%     \item \textit{Test and debug issues in a large code base.}
%     \item \textit{Understand how to conduct and respond to code reviews for proposed changes to a code base.}
%     \item \textit{Communicate how a code base works to others.}
%     \item \textit{Use documentation and Q\&A websites to learn just-in-time.}
%     \item \textit{Leverage Large Language Model tools for program comprehension and modification.}
% \end{itemize}

Throughout the course, students work on an open-source code base---\texttt{idlelib}\footnote{\url{https://github.com/python/cpython/tree/3.13/Lib/idlelib/}}, part of the CPython repository. The \texttt{idlelib} code base implements IDLE---a simple IDE meant for first-time programmers that comes with every Python installation. The code base is written completely in Python and consists of roughly 23,500 lines of code across 128 files, including roughly 10,000 lines of testing code. Throughout the term, students complete four projects individually in which they are asked to 1) configure and build a development environment, 2) modify an existing feature in IDLE, 3) add a new feature to IDLE, and 4) add a new feature to IDLE using GitHub Copilot to assist with the development process. Students also complete a group project with three other students in which they propose and implement a significant feature addition to the code base. 

\subsection{Participants}

Although our results related to students' trust attitudes includes data from the 71 students who answered all three trust surveys in this study, the start-of-term survey we administered to understand student demographics was completed by only 60 of those 71 students. This occurred because of a 2-week period at the beginning of the term in which students can add or remove courses from their schedule, meaning that several students eventually joined the course after the start-of-term survey was collected. Unfortunately, we cannot obtain the demographics of the 11 additional students in our dataset, but in this section, we will report on the demographics of the 60 students who completed the start-of-term survey. 

\textbf{Gender:} 67\% of respondents self-identified with \textit{he/him/his} pronouns and 33\% self-identified with \textit{she/her/hers} pronouns.

\textbf{Race:} Four-fifths (80\%) of students self-identify as Asian or Asian-American, 10\% as Latine, and 5\% as Caucasian. The remaining 5\% self-identified as Middle Eastern, Asian and Caucasian, and Latine and Caucasian. 

\begin{figure}
    \centering
    \includegraphics[scale=0.27]{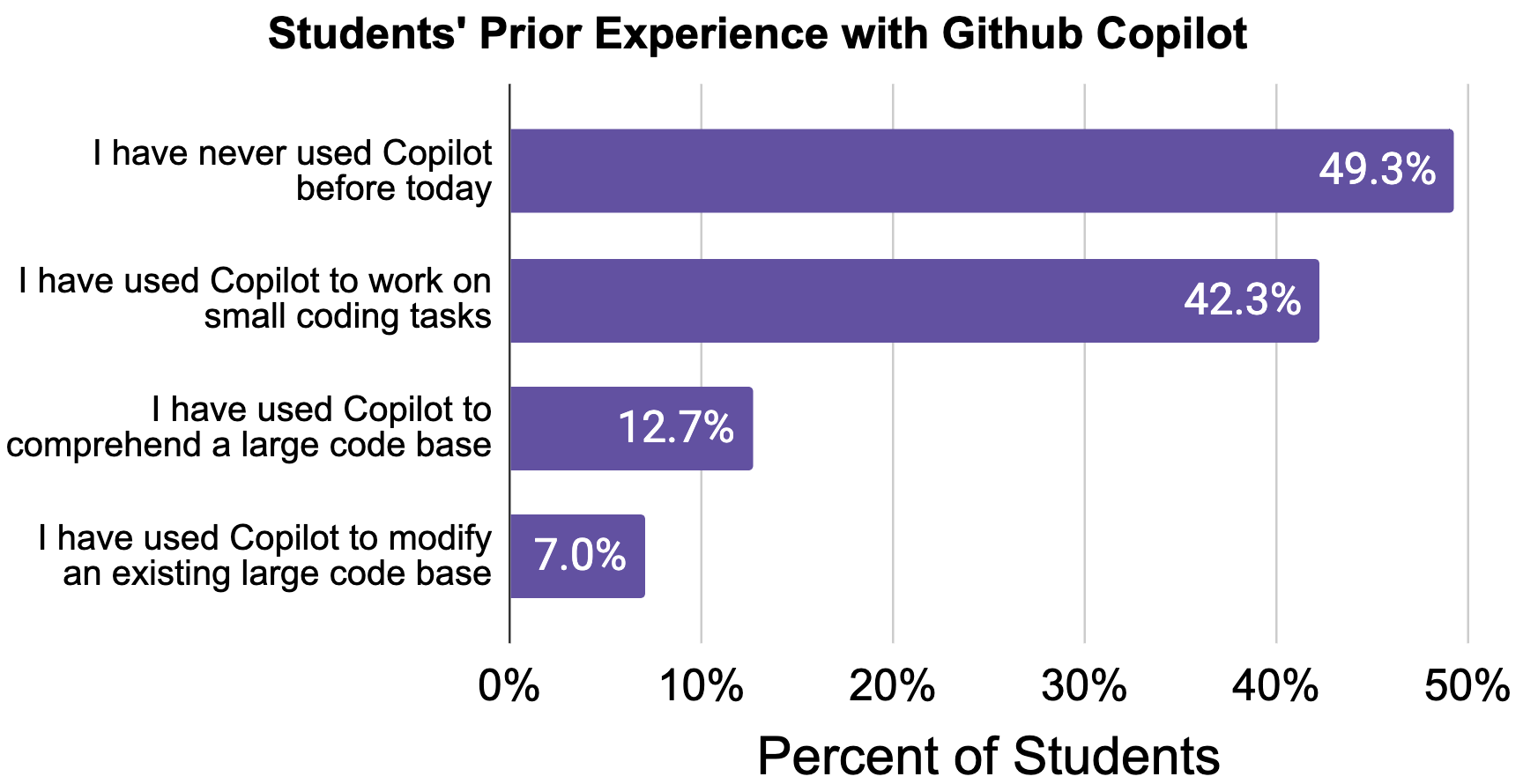}
    \caption{Student responses to the question: ``How much have you used GitHub Copilot before today? Check all that apply.''}
    \label{fig:copilot_experience}
\end{figure}

\textbf{Year-in-Program:} 47\% of students are fourth-year students, 37\% are third-year students, 13\% are second-year students, and 2\% are fifth-year students. The remaining students preferred not to disclose their year-in-program.

\textbf{Internship Experience:} Students self-reported their internship experience, with 57\% of students having \textit{no} internship experience, 28\% having completed 1 internship, and 15\% having completed two or more internships.

\textbf{First-Generation Student Status:} 45\% of students are first-generation students in their family, meaning that neither of their parents completed a bachelor's degree, whereas the remaining 55\% are not first-generation students.

\textbf{Native English Speakers:} Two-thirds (67\%) of students are native English speakers, meaning that their first language is English. 30\% are non-native English speakers and 3\% preferred not to answer.

\textbf{Prior GitHub Copilot Experience.} We asked about students' prior experience with Copilot on the Pre-Survey, so we have data from all 71 students. Figure \ref{fig:copilot_experience} shows that roughly half (49.3\%) of the 71 students had never used GitHub Copilot before the Pre-Survey, whereas the majority of students with Copilot experience had primarily used it for small coding tasks (42.3\% of all students).

\subsection{Students' Programming Environment}

At the start of the course, the instructor recommended that students use Visual Studio Code (VS Code), since lecture demonstrations would be conducted with this IDE. Since the course began in April 2025, the students used VS Code version 1.99\footnote{\url{https://code.visualstudio.com/updates/v1_99}}. Students were required to sign up for the GitHub Student Developer Pack \cite{GitHub_Education} so that they could access a free version of GitHub Copilot. To our knowledge, all students were able to access the Student Developer Pack by week 5 of the course. Students then installed the GitHub Copilot extension in VS Code \cite{Microsoft_2021}.

\section{Methods}
\label{sec:methods}

\subsection{Study Design}

\begin{figure*}
    \centering
    \includegraphics[scale=0.28]{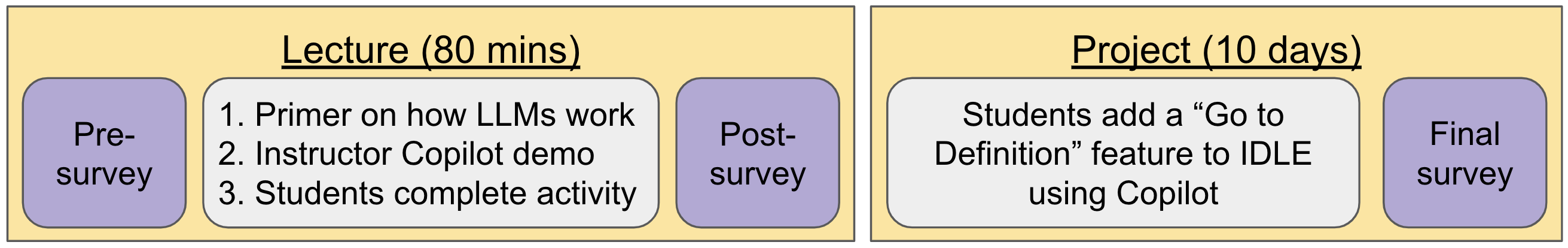}
    \caption{Our study design included a lecture and project component. Surveys were administered at the start and end of the lecture component and when students submitted their project.}
    \label{fig:design}
\end{figure*}

Figure \ref{fig:design} summarizes the two key components of our study. First, we introduced students to GitHub Copilot in one of the course lectures, which included a pre- and post-lecture survey about students' trust in GitHub Copilot. Second, students completed a project individually using GitHub Copilot to add a feature to IDLE.

\subsubsection{Lecture Component}

\begin{table}
    \centering
    \caption{Summary of the 80 minute lecture component.}
    \begin{tabular}{|C{1.5cm}|L{6cm}|}
    \toprule
    \hline
        \textbf{Time into Lecture} & \textbf{Activity} \\ \hline
        0:00-10:00 & Instructor made announcements about ongoing projects and course logistics. \\  \hline
        10:00-20:00 & Students completed the \textbf{Pre-Survey}. \\ \hline
        20:00-30:00 & Instructor explained the ``basics'' of NLP, including concepts like \textit{context}, \textit{training data}, \textit{probability distributions}, and \textit{non-determinism}. \\ \hline
        30:00-45:00 & Instructor introduced and demonstrated Copilot features, such as code generation, Copilot Chat, adding context, etc. \\ \hline
        45:00-58:00 & Students completed a code comprehension and modification task with Copilot. \\ \hline
        58:00-63:00 & Instructor demonstrated the solution to the task. \\ \hline 
        63:00-73:00 & Instructor demonstrated themselves using Copilot to complete a previous project in the course that students had already done.  \\ \hline
        73:00-80:00 & Students completed \textbf{Post-Survey}. \\ \hline
        \bottomrule
    \end{tabular}
    \label{tab:lecture_activities}
\end{table}

Table \ref{tab:lecture_activities} summarizes the 80 minute lecture component of our study. Following 10 minutes of announcements, students were given 10 minutes to complete a survey, which we will refer to as the \textit{Pre-Survey}. The Pre-Survey, which is described in Section \ref{sec:collection}, aimed to gather students' prior experiences using Copilot and initial trust level with Copilot before any influence from the instructor or the lecture. The lecture content itself included the basics of natural language processing and Copilot, demonstrations of key Copilot features, and an activity component in which students used Copilot to comprehend and modify a small feature in the code base. Notably, the instructor ensured they showed cases where Copilot was correct and incorrect during the demonstration. At multiple instances, the instructor pointed out when Copilot showed incorrect or non-optimal code output. The key features that the instructor demonstrated to students is listed below:

\begin{itemize}
    \item Code generation via function docstring comments.
    \item Code generation via in-line comments.
    \item Copilot chat.
    \item Adding the context of the entire code base or certain files and directories to prompts.
    \item Workspace commands (``/explain'', ``/fix'', ``/docs'', etc.).
\end{itemize}

Following the demonstrations and activities, students used the remaining 7 minutes of lecture time to complete the \textit{Post-Survey}. 

\subsubsection{Project Component}
After the lecture, students were assigned an individual project to implement a ``Go to Definition'' feature in IDLE \textit{using GitHub Copilot}. The ``Go to Definition'' feature refers to the code navigation shortcut that is typically included in powerful IDEs such as VS Code and Eclipse, allowing the user to select the ``Go to Definition'' option on a function call and be taken to the definition of that function. To correctly implement this feature, students need to 1) locate where the right-click menu view is defined in the code base, 2) add a ``Go to Definition'' option to the right-click menu, 3) implement the logic to identify the function definition, if it exists, when the option is invoked, 4) move the user's cursor to the function definition identified in Step 3 and 5) bind the ``Go to Definition'' menu option to the logic implemented in Steps 3 and 4. A solution to the task can be achieved by locating and understanding the right-click menu, creating one new Python file, and adding roughly 30 lines of code. Of course, student implementations may vary---some may not have added a new file and instead added their code to an existing file. Nonetheless, all students needed to both comprehend and generate code to implement the project. 

Following the project submission, students completed the \textit{Final Survey}, which asked them about their experience using Copilot for the individual project and their updated trust level. 

\subsection{Data Collection}
\label{sec:collection}
All three surveys had the same questionnaire, displayed in Table \ref{tab:trust_survey}, that was presented by \citeauthor{Amoozadeh_Trust} \cite{Amoozadeh_Trust}. In this section, we list the exact questions we asked in each of the three surveys.

\begin{table}
    \centering
    \caption{Survey on developer trust in AI systems from \citeauthor{Amoozadeh_Trust} \cite{Amoozadeh_Trust}, rephrased to use ``GitHub Copilot'' in place of ``AI system''. Students rate their agreement on a scale of 1 (Strong Disagree) to 5 (Strong Agree) for each statement.}
    \begin{tabular}{|C{0.8cm}|L{6.8cm}|}
    \toprule
    \hline
        \textbf{ID} & \textbf{Survey Question} \\ \hline
        S1 & I trust GitHub Copilot's output. \\  \hline
        S2 & The output GitHub Copilot produces is as good as that which a
highly competent person could produce. \\ \hline
        S3 & I know what will happen the next time I use GitHub Copilot
because I understand how it behaves. \\ \hline
        S4 & I believe the output of GitHub Copilot even when I don’t
know for certain that it is correct. \\ \hline
        S5 & I have a personal preference for using GitHub Copilot for my tasks. \\ \hline
        S6 & Overall, I trust GitHub Copilot. \\ \hline \bottomrule
    \end{tabular}
    
    \label{tab:trust_survey}
\end{table}

\subsubsection{Pre-Survey}

In addition to the 6-item trust survey, we asked:

\begin{itemize}
    \item How much have you used GitHub Copilot before today? Check all that apply. 
    %\item In a typical week, how often do you use Github Copilot?
    \item What factors or past experiences have contributed to your overall level of trust in GitHub Copilot for programming? 
\end{itemize}

\subsubsection{Post-Survey}
In addition to the 6-item trust survey, we asked:

\begin{itemize}
    \item Did you feel your overall trust in Copilot changed since the start of lecture? \textit{(My trust increased/No/My trust decreased)}
    \item Please explain your answer above. Why did your trust change (or why not)?
\end{itemize}

\subsubsection{Final Survey}
In addition to the 6-item trust survey, we asked:

\begin{itemize}
    \item Did you feel your overall trust in Copilot changed since the start of the project? \textit{(My trust increased/No/My trust decreased)}
    \item Please explain your answer above. Why did your trust change (or why not)?
\end{itemize}

\subsection{Data Analysis}

\subsubsection{Quantitative Analysis}

Our quantitative analysis involved two statistical tests. First, we conducted within-subjects, paired t-tests \cite{paired_ttest} for responses on the 6-item trust surveys. We chose to use a t-test given our sufficiently large sample size (n = 71) and since students recorded their answers before and after an intervention (whether it was attending lecture or completing the project). We conducted one round of paired t-test among the 6 pairs of Pre- and Post-Survey responses and an additional round of t-tests between the 6 pairs of Post-Survey and Final Survey responses. For each round of t-tests, we conducted Holm-Bonferroni corrections for multiple tests \cite{aickin_gensler_1996}. Second, we conducted a Chi-square test of trend \cite{chi_sq_trend} on student responses to the questions ``Did you feel your overall trust in Copilot changed since the start of the [lecture/project]?''  since their responses were categorical values with an order (e.g., ``My trust increased'' $\rightarrow$ ``No'' $\rightarrow$ ``My trust decreased''). This test looks for an association between the categorical values above and the condition, which in our case is immediate vs extended use of Copilot. 

\subsubsection{Qualitative Analysis}

Our qualitative analysis involved thematic analyses of the three open-ended survey questions discussed in Section \ref{sec:collection}. For all three questions, two researchers conducted a bottom-up approach to develop and revise a list of themes that were applied to each student's response (multiple themes could be applied to each response, if necessary). We used a negotiated agreement \cite{campbell2013negotiated} approach to resolve discrepancies between the two researchers. For the questions from the Post-Survey and Final Survey about students' change in trust in Copilot, \textit{71} student responses were labeled after narrowing the population to students who submitted the three surveys. However, we only analyzed the \textit{36} student responses to the Pre-Survey question about which past experiences impacted students' trust after narrowing the population to students who indicated that they had previously used Copilot on the Pre-Survey. We chose to exclude the answers from students who had no prior Copilot experience since these tended to either restate that they had no experience or they talked about other AI tools. For the three separate qualitative analyses that we conducted (one analysis for each open-ended question), the two researchers independently labeled a third of the student responses at a time, convened to discuss disagreements (i.e., negotiated agreement), reached a consensus, and revised the list of themes before labeling the next third of the responses.

\section{Results}
\label{sec:results}

\subsection{RQ1: How Trust Changes}

\subsubsection{Students' Overall ``Trust Paths''}

\begin{figure*}
    \centering
    \includegraphics[scale=0.4]{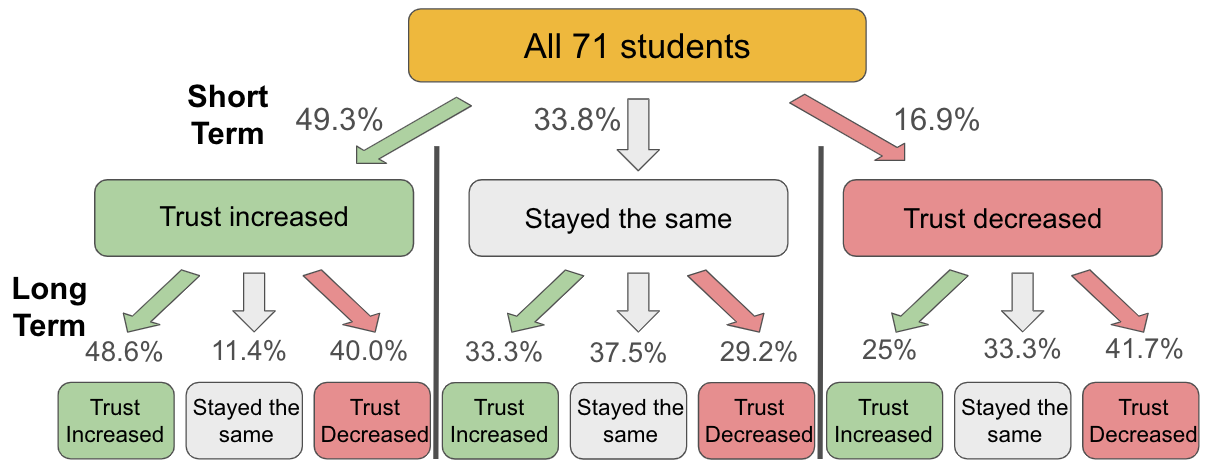}
    \caption{Students traveled through various ``trust paths'' after immediate and extended use. The ``immediate change'' layer has percentages that sum to 100. Each separated section in the ``extended change'' layer has percentages that sum to 100, since we display the proportion of students who chose each answer given their previous answer.}
    \label{fig:trust_path}
\end{figure*}

Figure \ref{fig:trust_path} shows the various ``trust paths'' that students exhibited via the two survey questions asking whether students felt their trust increased, decreased, or stayed the same. Nearly half (49.3\%) of the students increased their trust after the lecture component, with only 16.9\% having lower trust after the lecture. Of the students who had increased trust in the short term, however, these students seemed to either experience a clear increase (48.6\%) or decrease (40.0\%) in trust after the project component. The other two groups of students had a more even distribution of trust changes after extended use. 

\subsubsection{Changes Among Specific Trust Dimensions}

\begin{figure}
    \centering
    \includegraphics[scale=0.4]{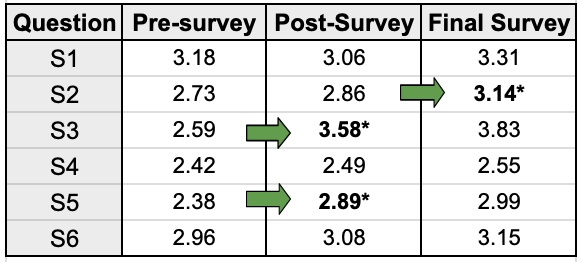}
    \caption{Student responses to the 6-item trust survey across all three surveys. The green arrow between two cells indicates a significant difference between those two survey items.}
    \label{fig:itemized_scores}
\end{figure}

Figure \ref{fig:itemized_scores} shows a breakdown of students' responses to the 6-item trust surveys, which provide a fine-grained analysis of what specific aspects of trust changed over time. Students' answers on the Pre-Survey were, on average, lower than 3.0 (the neutral rating), indicating that students tended to \textit{disagree} with the statements on the survey. Our paired t-test \cite{paired_ttest} and Holm-Bonferroni corrections revealed significant differences between two items on the Pre- and Post-Surveys and one item on the Post- and Final-Survey. Specifically, students had higher agreement with S3 (\textit{``I know what will happen the next time I use GitHub Copilot because I understand how it behaves''}) and S5 (\textit{``I have a personal preference for using GitHub Copilot for my tasks''}) in the immediate term. After extended use, students exhibited higher agreement to S2 (\textit{``The output GitHub Copilot produces is as good as that which a highly competent person could produce''}).

\subsubsection{Comparing Immediate vs Extended Trust Change}

Table \ref{fig:trust_change_comparison} shows the proportion of students who reported their trust increasing, staying the same, or decreasing after immediate and extended use. A Chi-square test of trend \cite{chi_sq_trend} shows that there is an significant association $(p = 0.03)$ between the time frame and the trust changes, with more students reporting trust increasing after immediate use and trust decreasing after extended use. 

% \begin{figure}
%     \centering
%     \includegraphics[scale=0.35]{figures/trust_change_comparison.png}
%     \caption{Comparison of trust changes after immediate and extended use. A Chi square test of trend shows a significant association between the time frame and students' responses.}
%     \label{fig:trust_change_comparison}
% \end{figure}

\begin{table}
\centering
\caption{Comparison of trust changes after immediate and extended use. A Chi square test of trend shows a significant association between the time frame and students' responses.}
\begin{tabular}{|l|c|c|c|}
\hline
 & \textbf{Increased} & \textbf{No Change} & \textbf{Decreased} \\
\hline
Immediate Use & 49.3\% & 33.8\% & 16.9\% \\ \hline
Extended Use & 39.4\% & 23.9\% & 36.6\% \\
\hline
\end{tabular}
\label{fig:trust_change_comparison}
\end{table}

\subsubsection{Impact of Prior Copilot Experience on Trust Changes}

Table \ref{fig:prior_exp_change} shows the breakdown of trust changes between students with and without prior Copilot experience. While a higher proportion of students without prior Copilot experience reported having increased trust in the immediate term than students with Copilot experience (60.0\% vs 38.9\%), a Chi-square test for trend revealed no significant differences in trust changes between students with and without prior Copilot experience.

\begin{table}
\centering
\caption{Comparison of change in trust level between students with and without prior Copilot experience, though no statistically significant difference is observed.}
\begin{tabular}{|l|c|c|c|}
\hline
 & \multicolumn{3}{c|}{\textbf{Immediate Use}} \\
\cline{2-4}
 & \textbf{Increased} & \textbf{No Change} & \textbf{Decreased} \\
\hline
Copilot Exp. & 38.9\% & 44.4\% & 16.7\% \\ 
No Copilot Exp. & 60.0\% & 22.9\% & 17.1\% \\
\hline
%\multicolumn{4}{|c|}{} \\
%\hline
 & \multicolumn{3}{c|}{\textbf{Extended Use}} \\
\cline{2-4}
 & \textbf{Increased} & \textbf{No Change} & \textbf{Decreased} \\
\hline
Copilot Exp. & 38.9\% & 22.2\% & 38.9\% \\
No Copilot Exp. & 40.0\% & 25.7\% & 34.3\% \\
\hline
\end{tabular}
\label{fig:prior_exp_change}
\end{table}

% \begin{figure}
%     \centering
%     \includegraphics[scale=0.4]{figures/prior_exp_trust_changed.png}
%     \caption{Comparison of change in trust level between students with and without prior Copilot experience, though no statistically significant difference is observed.}
%     \label{fig:prior_exp_change}
% \end{figure}

% \begin{figure*}
%     \centering
%     \includegraphics[scale=0.4]{figures/itemized_survey.png}
%     \caption{}
%     \label{fig:itemized_percent_changes}
% \end{figure*}

\subsection{RQ2: Why Trust Changes}

The full code books that we developed in our qualitative analysis is attached as supplemental material with this paper.

\subsubsection{Students' Initial Trust}

\begin{table}
    \centering
    \caption{Most common themes mentioned by students when asked about their prior experiences with Copilot (n = 36).}
    \begin{tabular}{L{5.5cm}|C{1.6cm}}
    \hline
    \textbf{Code} & \textbf{Frequency} \\ \hline
    Incorrect code output  &  28\% \\ \hline
    Uncertainty about Copilot response  &  19\% \\ \hline
    Correct code output  &  17\% \\ \hline
    Used Copilot for small tasks  &  14\% \\ \hline
    Prefers to use different LLMs  &  11\% \\ \hline
    LLM hallucinations  &  8\% \\ \hline
    Helped with productivity  &  8\% \\ \hline
    Copilot response not applicable to context  &  8\% \\ \hline
    Disruptive code suggestions  &  6\% \\ \hline
    \end{tabular}
    \label{tab:initial_perceptions}
\end{table}

Table \ref{tab:initial_perceptions} displays the most common themes mentioned by the 36 students who had prior experience with Copilot. Student responses tended to focus on Copilot's code generation abilities, mentioning incorrect (28\%) and correct code output (17\%), along with general uncertainty about Copilot's response (19\%), etc. Students also noted that LLMs are prone to hallucinate (8\%) and that Copilot's code suggestions were disruptive (6\%).

\subsubsection{Why Trust Changes after Immediate Use}

Tables \ref{tab:short_term_increase} and \ref{tab:short_term_decrease} show the most common reasons students gave for why their immediate trust increased or decreased. We do not display the reasons for trust staying the same in the immediate term due to space constraints. As Table \ref{tab:short_term_increase} shows, students attributed their increased trust to having learned about Copilot's features (29\%) that they were previously unaware of, seeing Copilot's ability to use the code base as context (26\%), and using it to support program comprehension in a large code base (20\%). Some students also expressed that they had a better understanding of how Copilot works ``behind-the-scenes'' (17\%), referring to the technical summary of natural language processing that the instructor presented in lecture.

\begin{table}
    \centering
    \caption{Most common themes mentioned by students whose trust increased after immediate use (n = 35).}
    \begin{tabular}{L{5.5cm}|C{1.6cm}}
    \hline
    \textbf{Code} & \textbf{Frequency} \\ \hline
    Learned about Copilot's features & 	29\%  \\ \hline
    Copilot's context awareness & 	26\%  \\ \hline
    Copilot's incorrect code output & 	26\%  \\ \hline
    Copilot's general correctness & 23\%  \\ \hline
    Copilot supports code comprehension & 	20\%  \\ \hline
    Understanding how Copilot works	 & 17\%  \\ \hline
    %Copilot depends on quality of prompt & 	11\%  \\ \hline
    % copilot's incorrectness & 	8.6\%  \\ \hline
    % copilot is not as good as a human & 	5.7\%  \\ \hline
    % copilot provides a scaffold & 	5.7\%  \\ \hline
    % industry adoption of copilot & 	2.9\%  \\ \hline
    % copilot is as good as a human & 	2.9\%  \\ \hline
    % copilot's poor code quality & 	2.9\%  \\ \hline
    \end{tabular}
    \label{tab:short_term_increase}
\end{table}

\begin{table}
    \centering
    \caption{Most common themes mentioned by students whose trust decreased after immediate use (n = 12).}
    \begin{tabular}{L{5.5cm}|C{1.6cm}}
    \hline
        \textbf{Code} & \textbf{Frequency} \\ \hline
  Copilot's incorrect code output & 	75\%  \\ \hline
Copilot's general incorrectness & 	25\%  \\ \hline
Copilot's variability	 & 17\%  \\ \hline
Copilot supports code comprehension & 	17\%  \\ \hline
% copilot provides a scaffold & 	8.3\%  \\ \hline
% copilot's correctness & 	8.3\%  \\ \hline
% copilot depends on quality of prompt & 	8.3\%  \\ \hline
% copilot's poor code quality & 	8.3\%  \\ \hline
    \end{tabular}
    \label{tab:short_term_decrease}
\end{table}

However, one of the most common responses given by students regardless of their trust change was related to Copilot's incorrect code output (26\% in Table \ref{tab:short_term_increase} and 75\% in Table \ref{tab:short_term_decrease}) or to Copilot's general incorrectness (25\% in Table \ref{tab:short_term_decrease}). Students whose trust decreased also pointed to Copilot's variability (17\%), or non-determinism, as a reason their trust decreased, which also was a theme introduced in the lecture component.

\subsubsection{Why Trust Changes after Extended Use}

Compared to their reasons for immediate trust changes, students had markedly different reasons for their trust changing after extended use. Tables \ref{tab:long_term_increase}, \ref{tab:long_term_constant}, and \ref{tab:long_term_decrease} highlight the common reasons for increased, constant, and decreased trust after extended use. Students seemed to have a split opinion on Copilot's code writing ability, with 57\% of students whose trust increased mentioning Copilot's correct code output (Table \ref{tab:long_term_increase}) but 54\% of students whose trust decreased mentioning Copilot's incorrect code output (Table \ref{tab:long_term_decrease}). Interestingly, 29\% of students whose trust stayed the same mentioned that \textit{Copilot performed as expected} (Table \ref{tab:long_term_constant}), suggesting they compared Copilot's outputs to their pre-conceived expectations of its performance.

\begin{table}
    \centering
    \caption{Most common themes mentioned by students whose trust increased after extended use (n = 28).}
    \begin{tabular}{L{5.5cm}|C{1.6cm}}
    \hline
    \textbf{Code} & \textbf{Frequency} \\ \hline
   Copilot's correct code output  &  57\% \\ \hline
Copilot requires a competent programmer  &  29\% \\ \hline
Copilot's general correctness  &  18\% \\ \hline
Copilot depends on the quality of prompts  &  18\% \\ \hline
Copilot provides a scaffold  &  14\% \\ \hline
Copilot supports code comprehension  &  14\% \\ \hline
    \end{tabular}
    \label{tab:long_term_increase}
\end{table}

\begin{table}
    \centering
    \caption{Most common themes mentioned by students whose trust stayed the same after extended use (n = 17).}
    \begin{tabular}{L{5.5cm}|C{1.6cm}}
    \hline
    \textbf{Code} & \textbf{Frequency} \\ \hline
Copilot performed as expected  &  29\% \\ \hline
Copilot requires a competent programmer  &  29\% \\ \hline
Copilot depends on the quality of prompts  &  24\% \\ \hline
    \end{tabular}
    \label{tab:long_term_constant}
\end{table}

\begin{table}
    \centering
    \caption{Most common themes mentioned by students whose trust decreased after extended use (n = 26).}
    \begin{tabular}{L{5.5cm}|C{1.6cm}}
    \hline
    \textbf{Code} & \textbf{Frequency} \\ \hline
Copilot requires a competent programmer  &  54\% \\ \hline
Copilot's incorrect code output  &  54\% \\ \hline
Copilot's inability to locate the correct code  &  31\% \\ \hline
Copilot's general incorrectness  &  23\% \\ \hline
Copilot's inability to debug its own code  &  12\% \\ \hline
    \end{tabular}
    \label{tab:long_term_decrease}
\end{table}

Two themes that emerged after extended use of Copilot were that \textit{Copilot requires a competent programmer} (36.6\% of all responses) and that \textit{Copilot depends on the quality of prompts} (15.5\% of all responses). The \textit{Copilot requires a competent programmer} label was applied when students mentioned needing to manually complete a part of the task that Copilot was unable to do. In a similar vein, the \textit{Copilot depends on the quality of prompts} was applied when students mentioned how they needed to modify their prompts to achieve the desired output. Both these themes relate to the programmer taking a more active role in the development process when Copilot is unable to deliver correct output.

\section{Discussion}
\label{sec:discussion}

\subsection{Interpretation of Results}

In this section, we aim to synthesize our many results to understand 1) students' initial attitudes and experiences with Copilot, 2) how and why trust changed after immediate use, and 3) how and why trust changed after extended use. 

\subsubsection{Students' Initial Perceptions and Experiences}
 
Our study involved a sample of 71 upper-division students who are about to graduate from university and enter the workforce. Their responses to the Pre-Survey provide a snapshot of their attitudes of and experiences with AI assistants \textit{before} they received any formal instruction or assignment requiring them to interact with such assistants. Our findings show that nearly half of all students had never used Copilot before our intervention, with 87\% of students having no experience with Copilot while working on large code bases. The 36 students who had any prior experience with Copilot expressed uncertainty in its output or reflected on the correctness or incorrectness of its output, indicating a general weariness of Copilot's output. Therefore, our findings suggest that before students learned about and used Copilot in the lecture, they lacked meaningful experiences with GitHub Copilot for authentic software engineering tasks.

\subsubsection{Immediate Trust Changes}

Following the lecture component, which is summarized in Table \ref{tab:lecture_activities}, the plurality (49.3\%) of students self-reported that their trust increased whereas roughly one-sixth (16.9\%) of students reported their trust decreasing. Taking a deeper look into this breakdown, 60.0\% of students \textit{without} Copilot experience reported that their trust increased, compared to only 38.9\% of those with Copilot experience reporting the same. Though the difference was not statistically significant, it may be the case that students are more sensitive to changes in trust at the beginning of their interactions with Copilot. If this is the case, CS educators should be mindful that how we present and discuss AI assistants can impact how students perceive such assistants, which aligns with previous work related to priming human's trust \cite{Pataranutaporn_Priming}.

We can also identify potential relationships between specific lecture activities and students' trust perceptions. For example, the instructor spent roughly 10 minutes in lecture discussing the basics of natural language processing and how Copilot actually makes predictions for any text or code output. Subsequently, students increased their agreement with item S3 that reads \textit{``I know what will happen the next time I use GitHub Copilot because I understand how it behaves.''} Therefore, we infer that our explicit instruction on how GitHub Copilot uses natural language processing to generate output helped shape students' expectations and understanding of how Copilot should behave.

Another aspect of the lecture focused on demonstrating Copilot's features, such as Copilot chat and adding context awareness (via either the ``@workspace'' tag or by using the ``Add context...'' option). Students also had time to practice using these features to complete a simple activity during lecture. Subsequently, students specifically mentioned that they \textit{Learned about Copilot's features} and \textit{Copilot's context awareness} as reasons for their increased trust after immediate use. As one student mentioned: \textit{``I never knew all of the tools you could use with Github Copilot in VScode. I feel like learning how to use these tools helped me understand how to use it more efficiently.}'' Therefore, we infer that making students aware of helpful Copilot features for working with a large code base contributed to higher student trust in Copilot after immediate use.

\subsubsection{Extended Trust Changes}

A remarkable proportion (36.6\%) of students mentioned that they felt \textit{Copilot requires a competent programmer} to use it. In this theme, students mentioned that they needed to manually find, comprehend, debug, or generate code due to Copilot's inability to do that subtask. For example, one student wrote that ``Despite adding multiple files and even the entire code base as context, it couldn't identify the right file to add the menu item to the right click menu. \textit{I had to go and find it myself after attempting to ask it for the solution many times.}'' Another noted that ``It was able to design the feature itself fine, but it can't handle the edge cases... \textit{I would usually have to be the one to tell Copilot what was wrong and how to possibly fix it.}'' These responses highlight the importance of programmers having the ability to not only recognize when Copilot is providing incorrect output, but also to comprehend or debug code without AI assistants for cases when the assistants are incorrect.

Despite this sentiment, we still saw a statistically significant increase in students' agreement with the statement \textit{``The output GitHub Copilot produces is as good as
that which a highly competent person could
produce''} on the trust survey. It seems counter-intuitive that students could have both seen Copilot generate output on par with a competent person but then also feel that Copilot requires a competent programmer to use it. However, we reason that the project component of the study involved a more substantial program comprehension and code generation effort than the simple activity in the lecture component. As a result, students must have observed some moments where Copilot completed a relatively challenging comprehension or generation task correctly \textit{and} some moments where it could not generate correct output, leading the programmer to manually write or comprehend part of the code. As a result, students' direct experience with using Copilot on challenging tasks helped students understand the abilities and, importantly, limitations of Copilot.

\subsection{Pedagogical Recommendations}

Our findings validate that student trust is dynamic. In both time frames we analyzed (immediate vs. extended use), student trust changed, whether measured by students’ responses to the 6-item survey or their self-assessed trust change. Since trust can change as students learn, use, and experiment with AI assistants, we make four concrete recommendations for CS educators to mediate student attitudes towards AI assistants. 

\subsubsection{Observation 1: Students had different reactions to the lecture and project components of the study.}

Our results showed a significant difference in how student trust changed after immediate and extended use (Table \ref{fig:trust_change_comparison}), with more students reporting that their trust decreased after extended use. We reason that since the tasks in the lecture component that shaped students' immediate trust were smaller in scale than the task in the project, students encountered more instances of Copilot's shortcomings in the project component. In fact, the project component resulted in a new theme mentioned by students that Copilot requires a competent programmer to manually complete parts of a task, indicating that students encountered instances when Copilot was unable to correctly do a task. 

\fbox{\parbox{0.93\linewidth}{\textbf{Recommendation 1}: In order to help students calibrate their trust and expectations of AI assistants, CS educators should provide opportunities for students to use AI programming assistants for tasks with a range of difficulty, including tasks within large code bases.}}

\subsubsection{Observation 2: Students valued Copilot's correctness (or incorrectness) on various tasks.}

Students frequently discussed Copilot's correctness or incorrectness on code writing, code comprehension, and debugging tasks as reasons for their trust changing. In terms of the MATCH model, these students were making \textit{trust judgments} based on the correctness of Copilot's output, implying a certain level of either systematic or heuristic processing. However, in order for students to make the \textit{correct} trust judgment, they must be able to verify Copilot's output across any task that they may use Copilot for. Though the previous recommendation highlights the importance of students having the chance to work with AI assistants, it is still vital for students to be able to verify the AI-generated output.

\fbox{\parbox{0.93\linewidth}{\textbf{Recommendation 2}: In order to support students' trust judgments of AI assistants' output, CS educators should ensure that students can still comprehend, modify, debug, and test code in large code bases \textit{without} AI assistants.}}

\subsubsection{Observation 3: Students valued learning how Copilot works.}

After the lecture component, some students felt a higher level of trust due to \textit{Understanding how Copilot works}. In fact, students had a statistically significant increase in agreement with item S3 on the trust survey that relates to understanding how a system behaves, which includes a system's process of generating output. By understanding how Copilot uses training data and context to determine the likelihood of the following token, students gained an understanding of the \textit{process integrity} that GitHub Copilot uses to generate responses. Such an understanding is important for students to recognize Copilot's abilities and shortcomings, allowing them to temper their expectations for Copilot's behavior.

\fbox{\parbox{0.93\linewidth}{\textbf{Recommendation 3}: CS educators should ensure that students are aware of \textit{how} AI assistants generate output via natural language processing so that students understand the AI assistants' expected behavior. }}

\subsubsection{Observation 4: Students valued learning about Copilot's features.}

Numerous students expressed the value of specific Copilot features, such as Copilot Chat, the ability to add context to Copilot Chat prompts, and keyword options such as ``/explain'', ``/fix'', and ``/docs.'' Based on our qualitative analysis, these interaction affordances---ways in which users can interact with Copilot---seemed to help students develop trust with Copilot. Since students began the study with limited experience using Copilot to work with large code bases, they were likely unfamiliar with Copilot's features that specifically support development in large code bases. However, the lecture and project components exposed students to Copilot's features and abilities in an authentic context, which the professional developers in \citeauthor{Cheng_OnlineCommunities}'s study noted as an important factor that impacts their trust and adoption of a tool \cite{Cheng_OnlineCommunities}.

\fbox{\parbox{0.93\linewidth}{\textbf{Recommendation 4}: CS educators should explicitly inform and demonstrate key features of AI assistants that are useful for contributing to a large code base, such as adding files as context and using keywords such as ``/explain'' and ``/docs'' in GitHub Copilot.}}

\subsection{Limitations}

Our study is subject to several limitations. One limitation is that we only investigate students' trust as an \textit{attitude} (via surveys and open-ended questions) rather than as a \textit{behavior} (via acceptance rate). Though trust plays an important role in programmers' use of a tool, our study stops short of investigating the impact of trust on programming behaviors. Further, our findings are limited to students with similar programming backgrounds and experience levels with Copilot, as well as the specific context of working with a Python code base with roughly 23,000 lines of code. Programmers with various levels of experience that are working in different code bases may develop distinct trust attitudes than our population did. Finally, our study also examines students' trust over a span of roughly 10 days. Though this time frame is still an improvement over prior works that have measured trust at a single point in time, we cannot extrapolate our findings to a longer time frame.

We also highlight several threats to validity in our study. The very act of the instructor introducing Copilot to students and mentioning the adoption of Copilot in the software engineering industry may lead to students trusting Copilot more by applying the \textit{endorsement heuristic} \cite{Mena_Endorsement}. The tacit ``endorsement'' by the instructor and the software engineering industry may have skewed student responses on the trust survey. Our approach to measuring trust also introduces several threats to validity. Given our choice of using a survey to measure trust, students may have been primed to consider and reflect on their trust level in Copilot. This may have evoked \textit{priming effects} that can alter students' perceptions of Copilot \cite{Pataranutaporn_Priming}, providing an inaccurate representation of student trust.

\label{sec:limitations}

\section{Conclusion}
\label{sec:conclusion}

A student's ability to calibrate their trust with AI assistants is a vital component of their AI literacy. In this study, we have uncovered how and why upper-division CS students' trust in GitHub Copilot changes as they learn about and use the tool. We show that some aspects of student trust increased, on average, after immediate and extended use of Copilot, but students expressed concerns with Copilot's ability to reliably comprehend, generate, and debug code in a large code base. Our findings highlight the factors that influence students' trust, including the correctness (or incorrectness) of Copilot's output, an understanding of how Copilot uses natural language processing to generate output, and an awareness of Copilot's features and user options. However, a dominant theme that emerged after students completed a 10-day project was that Copilot required a competent programmer to manually carry out some parts of the project. Based on these findings, we make 4 recommendations to CS educators: 1) educators should provide students with opportunities to work with AI assistants for tasks with a range of difficulty in a large code base, 2) educators should still ensure that students can comprehend, modify, debug, and test code in large code bases \textit{without} AI assistants, 3) educators should explain how AI assistants use natural language processing, and 4) educators should explicitly introduce and demonstrate an AI assistant's available features so students can effectively leverage the tool.

\begin{acks}
This work was supported by NSF \#2417531.
\end{acks}

\bibliographystyle{ACM-Reference-Format}

\bibliography{citations}

\end{document}